\documentclass[]{spie}  

\usepackage{amsmath,amsfonts,amssymb}
\usepackage{graphicx}
\usepackage[colorlinks=true, allcolors=blue]{hyperref}

\title{Feasibility of up-the-ramp sampling under variable sky for ground-based spectrographs}

\author[a,b]{Gaia Gaspar}
\author[a]{Marcin Sawicki}
\author[c]{Nelson Nunes}
\author[d]{Rub\'en J. D\'iaz}
\author[d]{James E. H. Turner}
\affil[a]{Institute for Computational Astrophysics and Department of Astronomy and Physics, Saint Mary's University, 923 Robie Street, Halifax, NS B3H 3C3, Canada}
\affil[b]{Observatorio Astronómico de Córdoba, Universidad Nacional de Córdoba, Laprida 854, X5000, Córdoba, Argentina}
\affil[c]{York University, 4700 Keele St, Toronto, ON M3J 1P3, Canada}
\affil[d]{Gemini Observatory, NSF NOIRLab, La Serena, Chile}

\authorinfo{Further author information: (Send correspondence to G. Gaspar)\\ E-mail: gaiagaspar@gmail.com / gaia.gaspar@smu.ca}

\begin{document} 
\maketitle

\begin{abstract}
Many modern near-infrared instruments employ HAWAII-2RG (H2RG) detectors with integration times that can reach 300-600s. Up-the-ramp (UTR) sampling offers advantages over Fowler sampling, including superior cosmic ray rejection and noise reduction, but requires fitting linear ramps from 30-60 reads. Ground-based K-band sky brightness has been reported to vary by 3-10\% on timescales of minutes, potentially introducing systematic errors and compromising photometric accuracy. Additionally, UTR data formats involve higher-dimensional FITS files with larger file sizes impacting observatory operations. 

We present a feasibility study using the GIRMOS Data Simulator with high-fidelity flux budgets and empirical K-band sky variations estimated, for Mauna Kea, from Gemini-NIRI at 10-20s cadence. Using a Monte Carlo approach we assess whether linear ramp fitting remains viable under variable sky conditions, quantify SNRs and systematic biases, and report nightly data volume estimates. Our results show that, in the H-band, the advantages of the UTR readout hold for read-noise-limited targets placed in the inter-line regions, translating into 3-4\% savings in observing time. The K-band inter-line regions do not show significant SNR improvement and can even degrade it due to the dominance of shot-noise generated by the thermal emission of the instrument+telescope system. In these regions, cosmic ray rejection recovers $>$ 98\% of events with false positive rates below 0.1\%, even under high sky variability. Over the sky emission lines, UTR fitting remains possible but its performance is compromised, both by a degradation in SNR and by a high rate of pixels falsely flagged by the cosmic ray rejection algorithm under highly variable sky. These findings address how ground-based conditions affect UTR implementation in near-infrared spectrographs, with GIRMOS as a concrete case of study.

\end{abstract}


\section{INTRODUCTION}
\label{sec:intro}  

The UTR sampling technique consists of fitting the slope of cumulative non-destructive reads. Its power is threefold: (1) enhances CR detection and removal by identifying jumps in individual reads (2) helps mitigate saturation by allowing the dropping of all the subsequent reads after the saturation point (3) In the read-noise-dominated regime, i.e. when readout noise is higher than the shot noise from the signal, UTR sampling improves the SNR over correlated double sampling (CDS) by a factor that grows asymptotically as $\sqrt{N/6}$ for large N\cite{Robberto2007, Rauscher2007, Vacca_2004}, where N is the number of non-destructive reads. UTR also presents higher performance compared to Multi-CDS (MCDS) readouts, commonly referred to as Fowler readout modes algorithms \cite{Fowler1990}. The reader can see Refs. \cite{Kubik2015, Brandt2024, Wang2025} for details on the performance of different fitting algorithms.

UTR sampling has been successfully implemented in space based telescopes, where the background is expected to be stable in comparison to ground conditions (but see Ref. \cite{Brammer2016} for an analysis on the impact of variable Helium emission in the upper atmosphere). Starting with  the WFC3\cite{wfc3irstarterguide} instrument aboard the Hubble Space Telescope (HST) and following with NIRSpec\cite{Boker2022} on the James Webb Space Telescope (JWST), among many others, the era of ramp fitting keeps evolving. Ground-based observations pose a different challenge: both PSF and sky brightness can be highly variable potentially affecting the count rates of the final images and the jump detection algorithms.

Nevertheless, a few ground-based instruments have adopted the UTR readout. The Habitable-Zone Planet Finder (HPF) \cite{Ninan2018} at McDonald Observatory operates in the J-band with a 1.7 $\mu$m cutoff H2RG detector and employs a dedicated fiber for sky monitoring, with the subtraction of the variable sky from the slope image happening after ramp fitting rather than accounting for it during the fit. The Eris/SPIFFIER integral field spectrograph \cite{George2016} at VLT has recently started offering UTR sampling in its slow read mode for the JHK-bands, but no reports on the performance of this readout mode have been published yet. CHARIS at the Subaru Telescope performs UTR fitting in the JHK-bands on high-contrast, short-exposure IFS data \cite{Brandt2017}, where the short integration times make sky variability a minor concern. Despite the ongoing adoption of UTR sampling in ground-based infrared instrumentation, the impact of temporally variable sky emission on ramp fitting performance has not been systematically characterized.

To understand the impact of the UTR sampling in ground-based observations we need to characterize the sky variations in timescales of 10-20s, the typical time ranges for saved ramps. Early studies by Ref. \cite{Ramsay1992} established that typical variations in the flux of OH are of the order of 3-10\% in time ranges of 10-15 minutes with a sampling cadence of $\sim$ 2 min. This was observed in spectra covering the 1-2.5 $\mu$m regime taken at Mauna Kea using the CGS2 spectrometer at the United Kingdom Infrared Telescope (UKIRT).
At Lenghu, China, Li et al. (2024) \cite{Li2024} reported, for the J and H' filters, the sky drift over several hours after sunset or before sunrise, showing that the integrated sky brightness can change by roughly a magnitude in a time span of 5-6 hours.

In this work we will explore the performance of UTR sampling for medium resolution NIR spectrographs taking as a case of study the Gemini Infrared Multi-Object Spectrograph (GIRMOS). GIRMOS\cite{Sivanandam2024} is an adaptive-optics-fed, near-infrared, multi-object integral field spectrograph under development by a consortium of Canadian Universities. It has successfully passed the Critical Design Review (CDR) stage and first light is expected before the end of the decade. Its complex design includes an Imager ($85^{\prime\prime}\times85^{\prime\prime}$ FoV) and 4 Integral Field Units (IFUs) that will patrol the $120^{\prime\prime}$ circular field of regard. It is expected to be fed by the Gemini North Adaptive Optics system \cite{Sivo2022} (GNAO) AO facility.

GIRMOS will operate in the Near Infrared (NIR), including the broad band filters J, H, and K and the broader, lower resolution filters YJ, JH, and HK, in exposures of up to 300-600s, corresponding to resolving powers of R8000 and R3000 respectively. 
The design contemplates the usage of a HAWAII-4RG detector for the Imager and 4 HAWAII-2RG detectors for the IFUs. Teledyne’s HxRG detectors are widely used in both space and ground-based infrared astronomical instrumentation, which makes studies of their noise performances easily accessible, see for example Refs. \cite{Benford2008, Kubik2015 ,Rauscher2022,Wang2025}

The leading science case for GIRMOS consists of a survey of distant galaxies and therefore imposes stringent requirements on the SNR of the reduced data, which for a significant number of targets will consist of emission lines with noise-buried continuum. This means that, provided the targets are selected to place relevant emission lines in the dark inter-line regions of the sky emission spectrum, the data could lie in the read-noise-dominated regime where UTR sampling offers its greatest advantage over alternative readout modes \cite{Robberto2007}. 
The inter-line continuum at Mauna Kea has been measured to be faint at high resolving powers for the H-band \cite{Maihara1993, Sullivan2012}, but at the moderate resolving powers of GIRMOS (R3000, R8000) the inter-line background is expected to be higher due to contamination from unresolved wings of OH
lines \cite{Sullivan2012}.The K-band of the NIR presents the additional challenge of the rising thermal emission from the sky, the telescope, the AO instrument and GIRMOS itself. This background contributes shot noise and can vary over time. It is therefore of interest to determine how the transition between read-noise- and background-limited regimes impacts the UTR sampling for instruments like GIRMOS.  

In this work we present a simulation-based study to answer the question on the feasibility and quality of UTR fitting under realistic sky variations for the K (R8000) and HK (R3000) GIRMOS configurations. We report systematic bias assessments and SNR changes compared to a Fowler-type readout and cosmic ray rejection performance while fitting UTRs, under absent, stable and variable sky conditions.

\section{METHODS}

In this Section we describe the experimental setup and the simulated data used to assess the performance of UTR sampling under variable sky conditions. Subsection \ref{sec:NIRI} describes the sky variability measurements obtained from archival NIRI K-band data. Subsection \ref{sec:simulated_data} details the GIRMOS Data Simulator, including flux budgets, noise models, and detector features.  Subsection \ref{sec:exp_design} presents the experimental design, including the simulated readout modes, Monte Carlo methodology, and the selected GIRMOS configuration, chosen to represent the most susceptible case to sky variability. Subsection \ref{sec:regions} defines the spectral regions used for the analysis.

\subsection{Sky variability measurements}
\label{sec:NIRI}


To characterize K-band sky variability on timescales relevant to GIRMOS exposures, we analyzed archival imaging from NIRI at the Gemini North telescope on Mauna Kea (Program GN-2022A-DD-106). NIRI is an Altair AO-fed near-infrared imager operating in the J, H, and K-bands, making it a suitable proxy for GIRMOS-like observations. We selected a K-band sequence of the ZTF-GPT-1 star with individual frame cadences of 10–20 s, matching the expected cadence of GIRMOS saved ramps. Sky brightness was measured in three source-free apertures across the sequence.

\begin{figure} [ht]
\begin{center}
\begin{tabular}{c} 
\includegraphics[height=7cm]{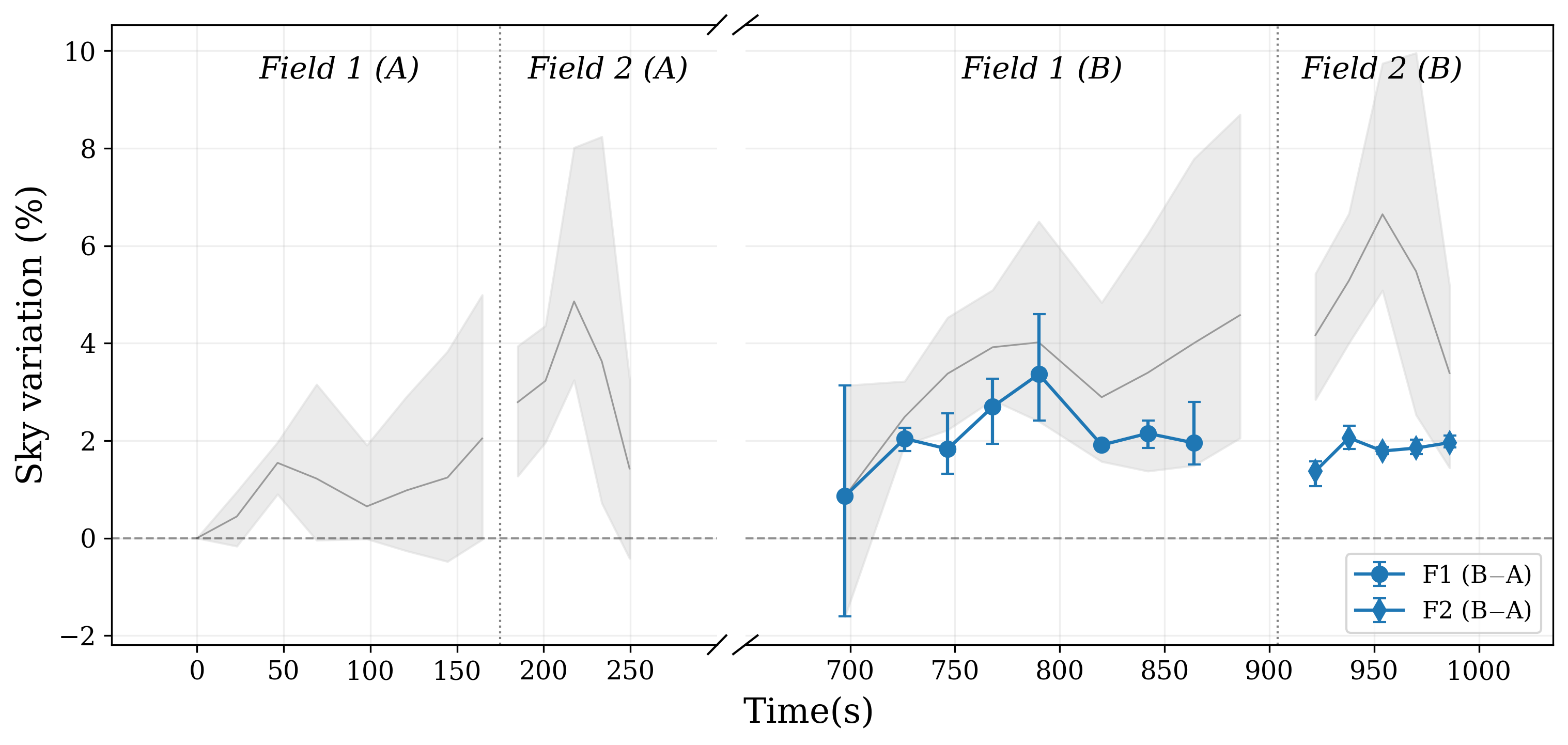}
\end{tabular}
\end{center}
\caption[example] 
{ \label{fig:fig_NIRI} 
K-band sky brightness variation measured from NIRI observations over $\sim$ 1000 s. Gray envelopes show the raw normalized measurements across three apertures per visit, including a dither-correlated spatial systematic. Colored points with error bars show the residual after subtracting matched dither pairs (B$-$A), revealing a smooth drift of 2–3\% over 1000 s. The broken x-axis omits a $\sim$ 400 s gap between visits A and B.}
\end{figure} 

As an AO-fed instrument, GIRMOS will operate under a restricted set of atmospheric conditions, with the worst-case limits being: IQ: 70\%, CC: 50\%, WV: any, SB: 80\% and airmass below 1.5. The NIRI data set considered in this work was observed under IQ:70\%, CC:50\%, WV:80\%, SB:80\%, AIRM: 1.04, presenting a mean AO seeing of $0.32^{\prime\prime}$ reported by Altair. Therefore, the dataset overall represents the typical-to-worst conditions under which GIRMOS will operate. A description of the observing conditions constraints defined by the Gemini Observatory can be found at the Observatory web page on Ref. \cite{gemini-sites}.

The observation spans $\sim$ 1000 s and targets two fields at different sky positions, each visited twice (subsets A and B) in the order F1A, F2A, F1B, F2B, with a ~400 s gap between visits A and B. Within each visit, the telescope follows a dither pattern of 8 (Field 1) and 5 (Field 2) pointings at ~20 s cadence. Crucially, subsets A and B share the same dither sequence, and we observe that the frame-to-frame brightness pattern within each field is nearly identical between visits — indicating a spatial systematic (likely flat-field residuals) that tracks the dither position rather than a true temporal signal. To isolate the genuine sky variability, we subtract each A frame from its corresponding B frame (matched by dither index), removing the position-dependent component. In Figure \ref{fig:fig_NIRI}, gray shaded envelopes show the min–max spread of the three apertures for the raw normalized measurements per subset, while colored points show the mean B$-$A difference (as a percentage of the first-frame brightness) with min–max error bars across the three apertures. The corrected time series, depicted in blue, reveals a smooth monotonic drift of 2–3\% over 1000 s, with no evidence of rapid transients at this cadence. For the simulations presented in this work, we conservatively adopt a sky variability model consisting of a linear drift of 2\% over 700 s with superimposed stochastic Gaussian pulses of up to 8\% peak amplitude, 20 s width, occurring at a rate of 18 per hour, yielding $\sim$3 peaks in 600 seconds. This peak amplitude is supported by two additional NIRI K-band sequences taken at an image quality poorer than the GIRMOS limit (IQ: 85\%, CC: 50\%, WV: 50\%, SB: 50\%, airmass 1.29 and 1.81), in which the normalized sky brightness shows oscillations of 8-10\% that rise and return to baseline within the $\sim$160-175 s covered by each sequence; we therefore consider 8\% a conservative upper bound for the pulse amplitude.

By applying the full observed variability amplitude uniformly to all spectral channels, we avoid assumptions about the differential behavior of OH emission lines and thermal continuum. This model is physically motivated for sky-line regions, where rapid fluctuation is expected, and conservative for the continuum, where it acts as a stress test; treating them separately would only sharpen the contrast we report. A more detailed model incorporating individual OH line changes, line-to-line variability, the correlated evolution of rotational-vibrational bands, and independent continuum behavior would be needed to assess the impact on derived science products such as emission line ratios, equivalent widths, or stellar population analysis, but is beyond the scope of this ramp-fitting performance study.

In all our simulations the PSF was assumed to be constant. However, measurements from the NIRI images show that the enclosed energy of a point source can vary by up to 35\%; the inclusion of PSF variations is left for future work.

\subsection{Simulated data}
\label{sec:simulated_data}

The \texttt{GIRMOS Data Simulator} (\texttt{DSIM}) is a project centered on simulating realistic, sequential, and complete data sets, aimed at testing the GIRMOS data reduction pipeline and support observation planning and early decisions. It is separated into two programmatic parts: \texttt{Cubesim} generates flux data cubes with high fidelity flux budgets estimated for GIRMOS and includes a module for AO-PSF generation. After \texttt{Cubesim}, \texttt{Detsim} projects the flux data cube into the detector plane, adds detector features and noises, simulates readout modes, and dither patterns with correct WCS. Accordingly, \texttt{DSIM} generates all the needed calibration files to feed the data reduction pipeline.

\texttt{Cubesim} models individual GIRMOS observations with a Python source-plane cube
simulator that propagates a source model through the instrument, atmosphere,
detector, and adaptive-optics point-spread function (PSF) to provide the input
scene for detector-output simulations. The source can be
specified analytically, for example as a point-source or S\'ersic spatial
profile with a Gaussian emission line, or from file-based spatial and spectral
templates. The simulator constructs the source cube on the GIRMOS spatial and
wavelength grid, convolves the spatial profile with the supplied PSF, applies
atmospheric transmission, and converts target radiance to detected electrons
using the telescope aperture, optical throughput, spectral sampling, pixel
solid angle, and H2RG quantum-efficiency curve. The noise model includes target
Poisson noise, sky background, thermal background from warm optical surfaces,
dark current, and read noise over the requested exposure time and number of
exposures. For this work, only the noiseless flux cube from \texttt{Cubesim} was used; noise was injected entirely by DSIM.

The GIRMOS performance-budget framework and the updated combined GNAO+GIRMOS
imaging performance model are described by Ref.\
\cite{girmos-performance-budget,lamb2024gnao-girmos-imaging}. The numerical
throughput, emissivity, and H2RG quantum-efficiency used here were provided by the GIRMOS project team through private communication \cite{girmos-performance-budget-private}.
The atmospheric transmission and sky-emission spectra are the Mauna Kea
near-infrared tables distributed for the Gemini Integration Time Calculators
\cite{gemini-sites}. The transmission spectra were generated with the ATRAN
atmospheric-transmission model \cite{lord1992atran}. The sky-background tables
represent the Gemini ITC model sky emission; the dominant 1--2.5~$\mu$m OH
airglow component at Mauna Kea is described by Ref. \cite{Ramsay1992}. The PSFs are simulated with a hybrid approach that
interpolates over a grid of PSFs generated with OOMAO \cite{conan2014oomao}
and adds residual jitter estimated from TIPTOP simulations
\cite{neichel2020tiptop}, with both components configured for MOAO using the
GNAO/GIRMOS adaptive-optics configuration described by Refs.
\cite{conod2023girmos-moao,conod2024girmos-moao}.

\texttt{Detsim} models the detector signal chain following the physical sequence of charge accumulation in an H2RG infrared array. The noiseless source signal is first modulated by a sensitivity map comprising per-pixel quantum efficiency variations (drawn from a Gaussian with configurable $\sigma$) and a polynomial illumination pattern representing the optical path non-uniformity. For multi-read readout, the modulated signal is accumulated linearly over the requested number of groups; when sky variability is enabled, the sky contribution is integrated per-group with a time-dependent modulation factor derived from the NIRI measurements described in Subsection~\ref{sec:NIRI}.

Noise is injected following the physical order of signal formation. Shot noise is drawn from a Poisson distribution at each group, dark current is accumulated at a nominal rate of 0.15~e$^-$/s assumed from early laboratory measurements of the GIRMOS detectors at 78~K (private communication), and a population of hot pixels (0.01\% of the array at $10^4\times$ the nominal dark rate) is included as a fixed spatial pattern. Dead pixels are zeroed prior to read noise injection. Read noise is drawn per amplifier and per read from a Gaussian whose $\sigma$ is derived from GSAOI commissioning characterization \cite{Carrasco2012}; for the slow pixel clock (100~kHz) used in this work, the effective per-read noise corresponds to $\sigma_{\rm VF}\sqrt{N_{\rm Fowler}/2}$ with $N_{\rm Fowler}=8$, where $\sigma_{\rm VF}$ is the "Very Faint Objects" read noise GSAOI tier.

UTR sampled frames follow a MACC pattern \cite{Kubik2015}, parametrized by the number of averaged reads per saved-to-disk group (UTRFRAME), with the possibility of dropping reads using the UTRSKIP parameter. The amount of ramps saved to disk during an integration is parametrized by NGROUPS. Finally, a full exposure can contain several integrations separated by detector resets, which is governed by the NINT parameter. 

For MCDS-8, the simulator produces a single 2D frame whose effective integration time is reduced by the Fowler sampling overhead ($8 \times t_{\rm read}$) following Ref.\cite{Robberto2007}; the paired reads are averaged and subtracted to emulate the on-instrument processing. All noise sources use deterministic seeds derived from configuration parameters, ensuring reproducibility while maintaining statistical independence between groups, amplifiers, and noise/CR realizations. The signal remains in electrons throughout this work; the per-amplifier gain conversion to ADU available in DSIM was not applied. To model CRs we built two empirical libraries using GSAOI dark frames, the selected CR candidates were filtered with a shape criterion and two different energy cuts were implemented to filter low energy non-CR artifacts: 211 e- (from RTN decontamination) and 1000 e- matching an estimated CR hit rate following Ref. \cite{Groom2004}, yielding 51,329 and 3,468 events respectively. More details about these libraries are presented in Sec. \ref{sec:cr_model}.

\begin{figure} [ht]
\begin{center}
\begin{tabular}{c} 
\includegraphics[height=12cm]{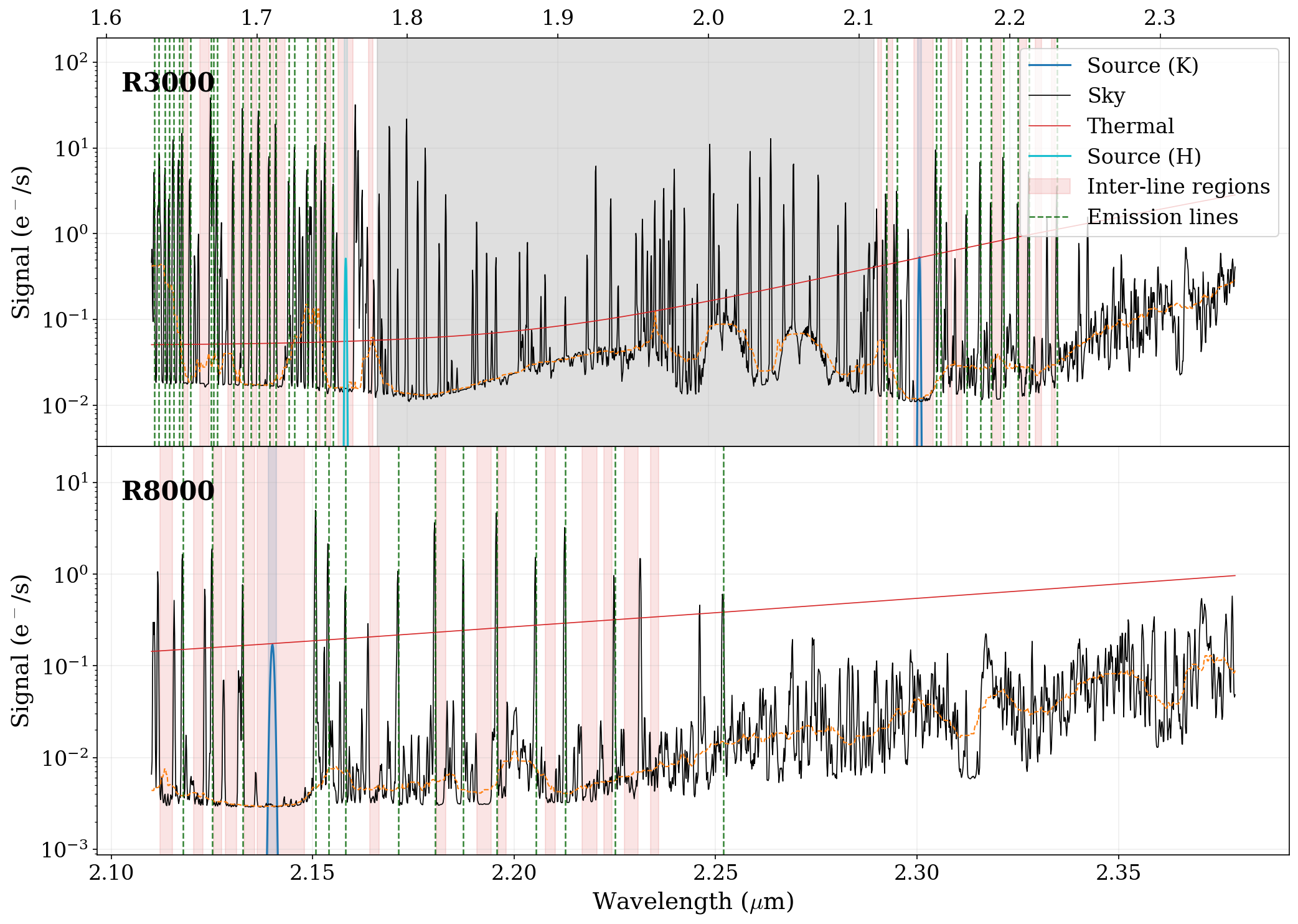}
\end{tabular}
\end{center}
\caption[example] 
{ \label{fig:frame_regions} 
Noiseless simulated spectra used in this work for the two available GIRMOS resolving powers: R3000 (top panel) and R8000 (bottom panel). The astronomical sources spectra consist of emission lines superimposed on a very faint continuum falling below the limits of the plot; the regions used to study each source are shaded in the same color as the emission line. The thermal spectrum, shown in red, includes the thermal emission from the combined telescope, adaptive optics (AO) system, and GIRMOS instrument. The spectral regions used in the analysis are indicated by pink shaded areas for the inter-line regions and by green dashed lines for pixels affected by sky emission lines. The dashed orange line shows the rolling median used to define the regions.}
\end{figure}

\subsection{Experimental design}
\label{sec:exp_design}

This experiment is designed to shed light on the question of fitting non-destructive reads under variable background, specifically variable sky emission at Mauna Kea. For this, we simulated realistic data frames considering two readout modes: UTRs and MCDS-8. MCDS-8 or Fowler-8 \cite{Fowler1990} is the most aggressive readout pattern implemented in the current Gemini instruments in terms of readout noise reduction. It is expected to improve the SNR by a factor of $\sim$ 2.2  with respect to the CDS readout under the readout-noise dominated regime (RON) \cite{Robberto2007}. Our goal is to compare the performance of UTR and MCDS-8 under three sky configurations: \texttt{no\_sky}, \texttt{stable} sky and \texttt{variable} sky.

For this purpose, \texttt{DSIM} (see Subsection \ref{sec:simulated_data}) was run in two steps. For the bias, SNR, and CR rejection analysis under the sky variability measured with NIRI data, 300 realizations were generated varying the seeds for the random noise sources (shot noise, readout noise, dark current, and CRs) for both the UTR and MCDS-8 readouts. The only difference being that the MCDS-8 simulations did not include CRs, as these are removed during a separate step in the data reduction pipeline for this readout mode. This choice slightly favors the MCDS-8 SNR, since CR-affected pixels are absent, but including them would introduce a bias in the opposite direction by contaminating the reference measurement. To further characterize CR rejection performance across a grid of sky variation levels, an additional 100 realizations were generated using the UTR readout only.

We included all the available detector features in \texttt{DSIM} to assess the performance of the ramp fitting under the most realistic case possible. This includes, along with the already mentioned random noises, QE sensitivity variations, per-slice illumination variations, and the thermal background of the combined system telescope + GNAO + GIRMOS. This thermal background was assumed to be constant during $\sim$600 s integrations.

GIRMOS will offer three options for plate scale: 0.1, 0.05, and 0.025 $^{\prime\prime}$/spaxel. Throughout this work, we adopt the 0.1 $^{\prime\prime}$/spaxel which subtends the largest solid angle per spaxel, producing the highest sky background per detector pixel and pushing the data further towards the shot-noise-dominated regime. Therefore, we consider this plate scale as the worst-case scenario for UTR performance. The two resolving powers offered for GIRMOS, R3000 (HK-band) and R8000 (K-band) are covered in this work.

For this experiment we assumed that the GIRMOS H2RG detector will be read non-destructively each t\_readout = 1.31 sec, following GSAOI slow readout mode \cite{Carrasco2012} (100 kHz), and using 32 readout channels, during an integration time of 620 seconds. For the UTR readout, dropping 7 reads (UTRSKIP=7) out of 10 produces 60 ramps saved to disk into a 4 dimensional data cube of shape (NINT, NGROUPS, NPIX, NPIX), where NINT is the number of integrations and NGROUPS is the amount of saved-to-disk ramps. NINT is set to 1 for this experiment. This gives a readout cadence for the saved reads of $\sim$10 seconds. The MCDS-8 mode is set to a slightly higher EXPTIME to match the integration time of UTR (see Ref. \cite{Robberto2007}) and the data format is one 2D image containing the multi-pair-averaged signal, emulating the on-instrument signal processing. 

For fitting the UTRs, the pure-Python-based module \texttt{fitramp} \cite{Brandt2024} is used. This code offers the elegance of fitting read differences instead of individual reads; this makes the covariance matrix semi-diagonal, allowing computation of quantities in closed form which leads to a computational cost that is linear in the number of resultants (reads or groups of on-instrument averaged reads). The code also enables jump-masking for CR rejection using a $\chi^2$ improvement approach \cite{Brandt2024b}.

We simulate a faint emission-line source representative of the GIRMOS high-redshift galaxy survey, with two emission lines placed in dark inter-line regions; one in the K-band at 2.14 $\mu$m and another in the H-band at 1.759 $\mu$m . The continuum is set at a level buried in the noise and the integrated flux of each line is 10$^{-16}$ erg s$^{-1}$ cm$^{-2}$, representative of the line fluxes for the targets planned for the GIRMOS survey. A comparable published survey can be found in Ref. \cite{Forster2018}. Figure \ref{fig:frame_regions} shows the count-rates of the noiseless simulated spectra separated into three components: the sky emission, thermal background from the instruments chain and the source spectrum.


\subsection{Region definitions}
\label{sec:regions}

The continuum inter-line regions for analysis were selected using a rolling median in 8 pixels for the R3000 mode and 12 pixels for the R8000 mode. Emission line regions were identified by visual inspection. Continuum region boundaries were adjusted by 1 pixel where they overlapped with emission line positions, to ensure mutually exclusive classification.

Beyond 2.25~$\mu$m the K-band becomes crowded with contiguous, relatively faint sky emission lines, making the identification of inter-line continuum regions impractical. This spectral region also receives the highest thermal background from the combined telescope and instrument optical path, and we consider it unsuitable for studies of faint extragalactic continuum sources.

For the statistical analysis, all inter-line continuum regions are combined into a single sample and all emission line regions into another; per-pixel statistics are computed over each combined sample.

The source regions in the two bands are computed using a simulation containing only the noiseless source, pixels above a threshold of 0.005 $e^-/s$ are selected and converted into a mask. This guaranties that all the values presented for these regions correspond to the core of the source, leaving out the wings of the lines. This prevents diluting the results by including faint pixels with low shot-noise, with the downside of regions containing only 155-171 pixels per frame (for R8000 and R3000 respectively), affecting the statistical significance of some results.

\section{RESULTS}

In this Section we describe two main results: Subsection \ref{sec:bias} shows the systematic bias and SNR changes involved in using UTR sampling under variable sky in comparison with the MCDS-8 baseline case. Subsection \ref{sec:rejection} describes the CR recovery and false positive rates obtained when fitting the ramp for both the baseline sky variation derived from NIRI data and a grid of increasing pulse amplitudes.

\subsection{Bias and SNR improvement}
\label{sec:bias}

To quantify the bias and SNR improvements when using the UTR readout across different sky modes we performed a Monte Carlo analysis with 300 noise realizations to determine per-pixel statistics. The simulations used in this work include all the noises/detector features available in \texttt{DSIM} (see Subsection \ref{sec:exp_design}). In principle, sky variations could cause systematic biases in the slope of the UTR fitting, which could translate into a bias in the count rates measured in UTR compared to other readout methods, directly affecting flux measurements. In parallel, sky variations could reduce or even reverse the SNR improvement of the UTR readout when applied under sky variations. 

\begin{figure} [ht]
\begin{center}
\begin{tabular}{c} 
\includegraphics[height=4.85cm]{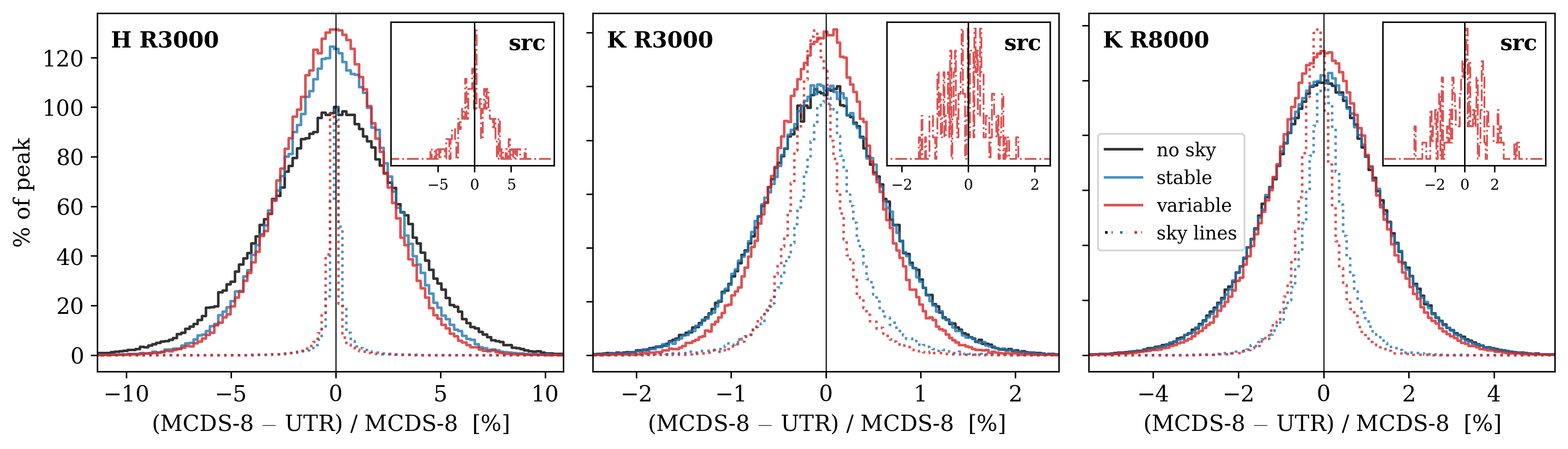}
\end{tabular}
\end{center}
\caption[example] 
{ \label{fig:bias} 
Per-pixel count-rate bias, in percent of the MCDS-8 value for the combined continuum regions (solid histograms), sky emission line regions (dotted line histograms) and source emission lines (top-right insets, only \texttt{variable} sky case). Values are averaged over 300 noise realizations.}
\end{figure}

To study the systematic bias we computed the mean count rate per-pixel for each readout mode across the 300 seeds, then expressed the difference (MCDS-8 $-$  UTR) as a percentage of the MCDS-8 count rate. Ramp differences were fitted using the two-iteration procedure recommended for \texttt{fitramp}, in which an initial slope estimate is used to rescale the covariance matrix before the final fit, effectively removing the estimator bias inherent to ramp fitting. Figure \ref{fig:bias} shows the distributions of these residuals for the two resolutions available and for three regions: the sum of the sky inter-line regions, the sum of the sky emission line regions and the region where the source emission lines were placed (see Subsection \ref{sec:regions} for more details). The distributions are shown for the 3 sky modes except for the source region where only the \texttt{variable} sky case is depicted.

\begin{figure} [ht]
\begin{center}
\begin{tabular}{c} 
\includegraphics[height=4.85cm]{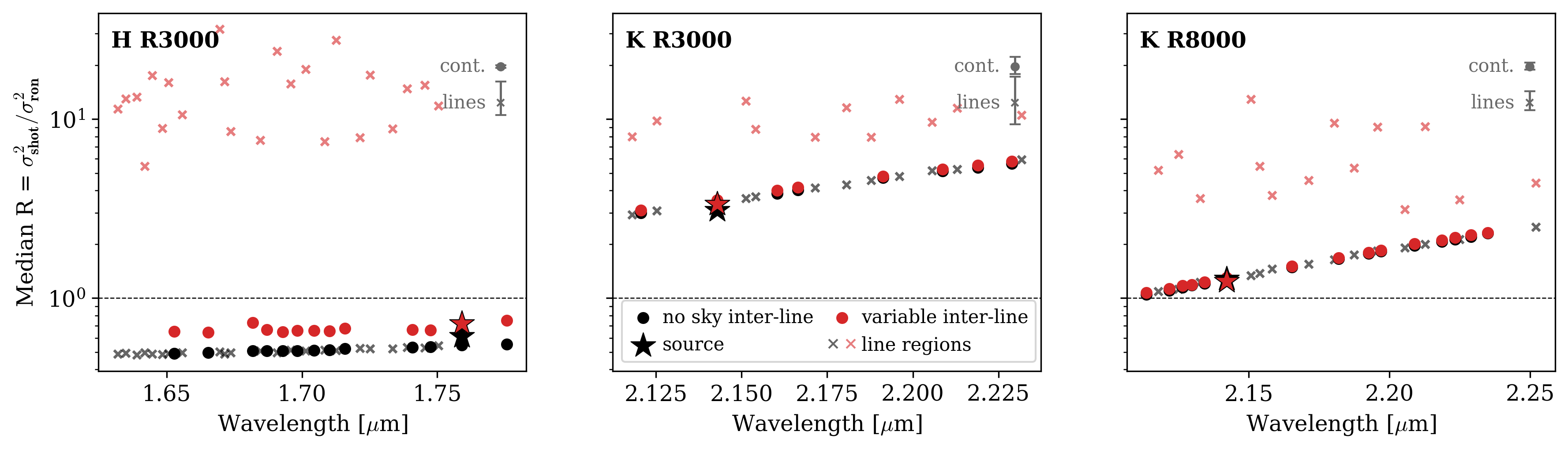}
\end{tabular}
\end{center}
\caption[example] 
{ \label{fig:noise_regimes} 
Region median per-pixel of the ratio between shot-noise and readout-noise. Grey error bars show the 16th–84th percentile range of each per-pixel distribution.}
\end{figure}

The largest systematic count-rate difference between UTR and MCDS-8, of 0.15\%, occurs in the sky line regions for the \texttt{variable} sky case at R3000, followed very closely by 0.14\% occurring in the H-R3000 source region for the \texttt{no\_sky} case. Overall, the \texttt{variable} sky mode shows biases that can reach 0.15\%, at least one of order of magnitude below the typical NIR flux uncertainties of 2-5\%. For \texttt{stable} sky, the bias always remain below 0.02\% indicating that sky variability has a quantifiable impact in the count rates with respect to MCDS-8 but, at least for the variability studied here, it is negligible in comparison to other uncertainties involved. The bias estimated for the sources has to be taken with caution as the number of pixels contained in those regions is small, causing noisy distributions (see insets in Figure \ref{fig:bias}) but is expected to remain below the estimated for the sky emission lines as they are less affected by shot-noise (see Figure \ref{fig:noise_regimes}) .

\begin{figure} [ht]
\begin{center}
\begin{tabular}{c} 
\includegraphics[height=4.85cm]{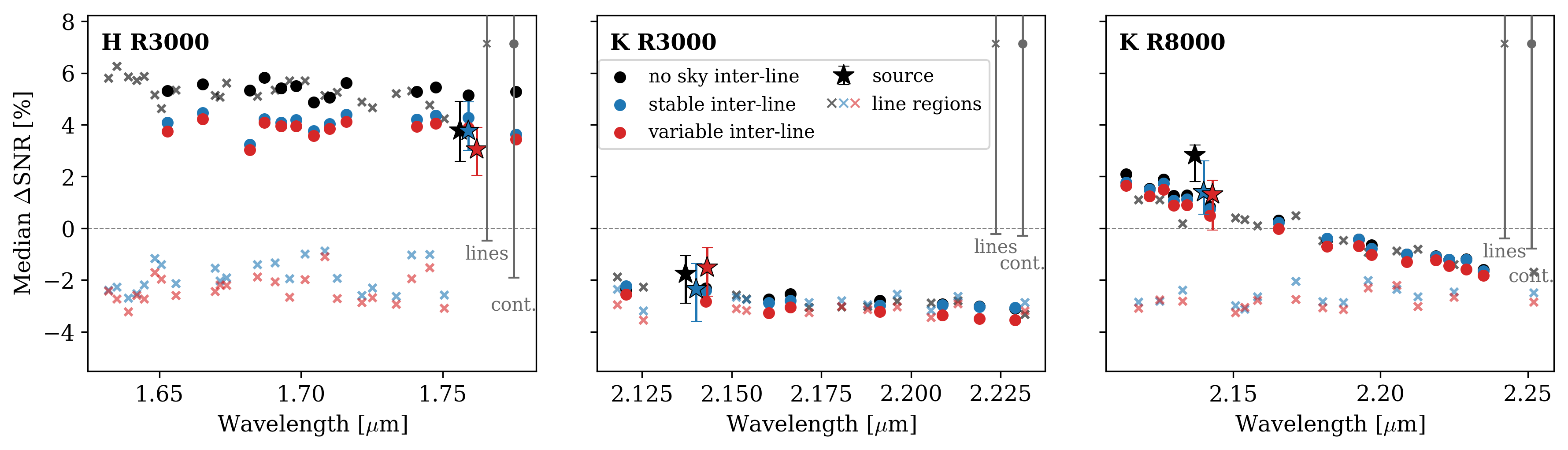}
\end{tabular}
\end{center}
\caption[example] 
{ \label{fig:snr_scatter} 
Per-pixel median SNR improvement between UTR and MCDS-8 readout modes; values above zero indicate that UTR improves the per-pixel SNR. Regions are displayed at their respective wavelengths rather than aggregated. Grey error bars show the 16th–84th percentile range of each per-pixel distribution. Source markers (stars) are artificially offset in wavelength for visibility; their error bars show the 16th–84th percentile of a bootstrap distribution (10,000 realisations of the median).}
\end{figure}

SNR variation across sky modes and resolutions is presented in Figure \ref{fig:snr_scatter}. The H-band shows a modest increase of SNR for UTR in the inter-line regions and for the source of 3-4\% even under variable sky, while sky lines present a degradation in the SNR in comparison to MCDS-8; this is driven by the dominating noise as shown in Figure \ref{fig:noise_regimes}, where the ratio between the variance produced by shot noise over readout noise is presented for the same regions. These regimes were calculated by simulating each noise separately. In the H-band both the inter-line regions and the source are dominated by readout noise, while the sky lines are in the shot-noise dominated regime.
For the K-band in the R3000 mode all the points lie in the shot-noise dominated regime, causing a degradation in the SNR of 2-4\% for UTR. In the K-band R8000 configuration a hybrid situation occurs: the bluest inter-line regions, including the source, show a 0-2\% improvement in SNR dropping after $\sim 2.15 \mu$m; the sky emission lines are always dominated by shot-noise and degraded by UTR. In the K-band inter-line regions, the dominant source of shot noise is the thermal spectrum, as shown by the \texttt{no\_sky} run, which contains only thermal emission in these regions and fully overlaps with the on-sky (\texttt{stable} and \texttt{variable}) runs, indicating that the sky continuum does not significantly contribute to the shot noise.

The SNR improvement in sources placed in the inter-line regions of the H-band can be translated into savings in observing time. For the read-noise-dominated targets, the effective read noise for a given readout mode is independent of integration time, so SNR scales linearly with exposure time: $\mathrm{SNR} = C \times T / \sigma_\mathrm{eff}$, where $C$ is the count rate, $T$ the exposure time, and $\sigma_\mathrm{eff}$ the effective read noise of the readout mode. To reach the same SNR with MCDS-8 as UTR delivers in time $T$, the required MCDS-8 exposure time is $T_\mathrm{MCDS} = (\sigma_\mathrm{MCDS}/\sigma_\mathrm{UTR}) \times T$. Since the bias between the two readout modes is negligible, the measured SNR ratio directly reflects the ratio of effective read noises: $\sigma_\mathrm{MCDS}/\sigma_\mathrm{UTR} = \mathrm{SNR_{UTR}}/\mathrm{SNR_{MCDS}}$. For R3000 H-band modes this translates to a 3-4\% saving in telescope time.

The 6\% SNR improvement measured for the H-band inter-line regions in the \texttt{no\_sky} run is consistent with the theoretical prediction for UTR with N=60 groups over MCDS-8 in the read-noise-limited regime, as derived from the formulae in Ref. \cite{Robberto2007} and consistent with previously reported values in Ref. \cite{Garnett1993}, confirming that our simulations recover the theoretical value in the absence of sky emission. As noted earlier, the 0.1 arcsec/pixel plate scale represents the highest background case for these filter/resolution configurations. We therefore expect the H-band at R8000 and the J-band configurations to yield improvements closer to the 6\% theoretical limit, as their lower sky and thermal background place them deeper into the read-noise-limited regime.


\subsection{CR rejection rates and false positives}
\label{sec:rejection}

One of the advantages of UTR sampling is its superior CR rejection, which relies on flagging outlier reads in the ramp, or "jumps", and discarding them from the slope fitting. A potential problem when fitting data from ground-based observations is the confusion of jumps, caused by sudden sky variations, with CRs, this could disrupt ramps, impacting the count rates in the final fitted image.

The fitting package chosen for this study, \texttt{fitramp}\cite{Brandt2024}, which fits read differences instead of single reads, uses a pair of thresholds to decide whether a read difference should be rejected or not \cite{Brandt2024b}. To do so, it analytically computes the $\chi^2$ values obtained when removing each read difference and compares with the given thresholds, the first one (oneomit) corresponds to the case of removing one difference and the second (twoomit) to the case of removing two differences in the same ramp. The default oneomit threshold corresponds to 4.5$\sigma$. 

Table \ref{tab:cr_rejection} presents the percentages of CRs successfully masked by \texttt{fitramp} compared to the CR maps produced by the data simulator and the percentages of false positives (FP) for all resolution/band/region studied. 

\begin{table}[ht]
\caption{CR rejection performance: mean per-pixel recovery and false positive rate evaluated on inter-line continuum, emission line, and source regions separately across 300 noise realizations. Uncertainties correspond to the standard deviations.}
\label{tab:cr_rejection}
\begin{center}
\begin{tabular}{|l|c|c|c|c|c|c|}
\hline
\rule[-1ex]{0pt}{3.5ex}  & \multicolumn{2}{c|}{Inter-line [\%]} & \multicolumn{2}{c|}{Emission lines [\%]} & \multicolumn{2}{c|}{Source [\%]} \\
\rule[-1ex]{0pt}{3.5ex} Sky mode & Rec. rate & FP & Rec. rate & FP & Rec. rate & FP \\
\hline
\multicolumn{7}{|c|}{\textbf{R8000 K-band (inter-line: 526,020, lines: 54,196, src: 155 px)}} \\
\hline
\rule[-1ex]{0pt}{3.5ex} No sky & $98.9 \pm 0.6$ & $0.08 \pm 0.00$ & $98.6 \pm 2$ & $0.09 \pm 0.01$ & $99.5 \pm 4$ & $0.05 \pm 0.2$ \\
\rule[-1ex]{0pt}{3.5ex} Stable & $98.8 \pm 0.6$ & $0.08 \pm 0.00$ & $97.7 \pm 3$ & $0.07 \pm 0.01$ & $99.5 \pm 4$ & $0.06 \pm 0.2$ \\
\rule[-1ex]{0pt}{3.5ex} Variable & $98.8 \pm 0.6$ & $0.08 \pm 0.00$ & $97.7 \pm 3$ & $0.09 \pm 0.01$ & $99.5 \pm 4$ & $0.07 \pm 0.2$ \\
\hline
\multicolumn{7}{|c|}{\textbf{R3000 H-band (inter-line: 204,032, lines: 114,768, src: 177 px)}} \\
\hline
\rule[-1ex]{0pt}{3.5ex} No sky & $99.0 \pm 0.9$ & $0.04 \pm 0.00$ & $99.0 \pm 0.9$ & $0.04 \pm 0.01$ & $99.5 \pm 4$ & $0.05 \pm 0.2$ \\
\rule[-1ex]{0pt}{3.5ex} Stable & $99.0 \pm 0.9$ & $0.04 \pm 0.01$ & $96.0 \pm 2$ & $0.05 \pm 0.01$ & $99.5 \pm 4$ & $0.03 \pm 0.1$ \\
\rule[-1ex]{0pt}{3.5ex} Variable & $99.0 \pm 0.9$ & $0.04 \pm 0.01$ & $96.0 \pm 2$ & $0.26 \pm 0.02$ & $99.3 \pm 5$ & $0.05 \pm 0.2$ \\
\hline
\multicolumn{7}{|c|}{\textbf{R3000 K-band (inter-line: 164,182, lines: 55,790, src: 171 px)}} \\
\hline
\rule[-1ex]{0pt}{3.5ex} No sky & $98.5 \pm 1$ & $0.09 \pm 0.01$ & $98.3 \pm 3$ & $0.08 \pm 0.01$ & $100.0 \pm 0.0$ & $0.08 \pm 0.3$ \\
\rule[-1ex]{0pt}{3.5ex} Stable & $98.5 \pm 2$ & $0.08 \pm 0.01$ & $96.9 \pm 3$ & $0.05 \pm 0.01$ & $98.6 \pm 12$ & $0.1 \pm 0.3$ \\
\rule[-1ex]{0pt}{3.5ex} Variable & $98.5 \pm 1$ & $0.08 \pm 0.01$ & $96.8 \pm 4$ & $0.08 \pm 0.01$ & $98.6 \pm 12$ & $0.07 \pm 0.2$ \\
\hline
\end{tabular}
\end{center}
\end{table}

CR recovery rates are presented only for the most complete CR library available in \texttt{DSIM}, implementing the RTN modeled cut at 211 $e^-$. Its energy distribution is shown in Figure~\ref{fig:cr_stamps}, and a detailed description of their construction is presented in Subsection \ref{sec:cr_model}.

For the continuum inter-line regions, the recovery rates for the K-band in both R8000 and R3000 modes are above 98\% and consistent across sky modes given the variances across seeds. For the H-band, the CR recovery rate rises to 99\% for all sky modes, consistent with the lower background in H-band, where CRs have slightly higher contrast against the noise floor. 

Over the sky emission line regions the CR recovery rates are lower, reaching 96\% in the worst case, occurring on the H-band R3000 mode, for the \texttt{stable} and \texttt{variable} modes. This consistency across on-sky modes indicates that the lower recovery rates are probably not due to variability but to shot-noise fluctuations in the high OH signal that can mask CR jumps. The K-band R3000 sky emission lines show a similar behavior, with rejection rates of $\sim$97\%. 

For the sources, the analysis of the CR rejection rates is less reliable due to the low pixel numbers involved; with only $\sim$ 170 source pixels per configuration, few CRs fall on the source region in any given realization, leading to large seed-to-seed variance that reaches $\sim$ 12\% in the worst case. The recovery rates are above 98.5\% and show no sky-mode dependence.

Some pixels are mistakenly flagged as CRs; the rates for each configuration are shown in the second column of each region in Table \ref{tab:cr_rejection}. Over the inter-line regions, for all filter/resolution configurations the FP rates are consistent across sky modes, with the H-band presenting half FP rate with respect to the K-band; shot-noise is very low in the H-band darker regions, so sporadic jumps caused by random higher realizations of shot-noise are rarer than in the thermal-dominated K-band. Over emission lines, for the K-band, the \texttt{variable} and \texttt{no\_sky} FP rates are consistent within uncertainties, while the \texttt{stable} case shows slightly lower rates. In H-band, the \texttt{variable} sky raises the FP rate to 0.25\%, significantly higher than the 0.04–0.05\% measured for the \texttt{no\_sky} and \texttt{stable} cases on the same regions, suggesting that at these high signal levels, sky jumps can cause false positives.

Finally, on the source regions, the FP is at maximum 0.11\% for the K-band R3000 stable sky case and is uniform across sky modes and configurations, with no evidence of variable sky causing a large number of false positives. The false positives over the sources are mostly flagged only in one seed, with only 5 pixels flagged more than once in the R3000 K-band source; none of them correspond to the peak of the source signal. No pixels are flagged more than once across 100 seeds for the H-band and the K-band R8000 sources. Overall, FP rates remain below 0.3\% across all regions and configurations studied. These results hold under the sky variations estimated for the weather conditions under which GIRMOS will operate.

\begin{figure} [ht]
\begin{center}
\begin{tabular}{c} 
\includegraphics[height=5.5cm]{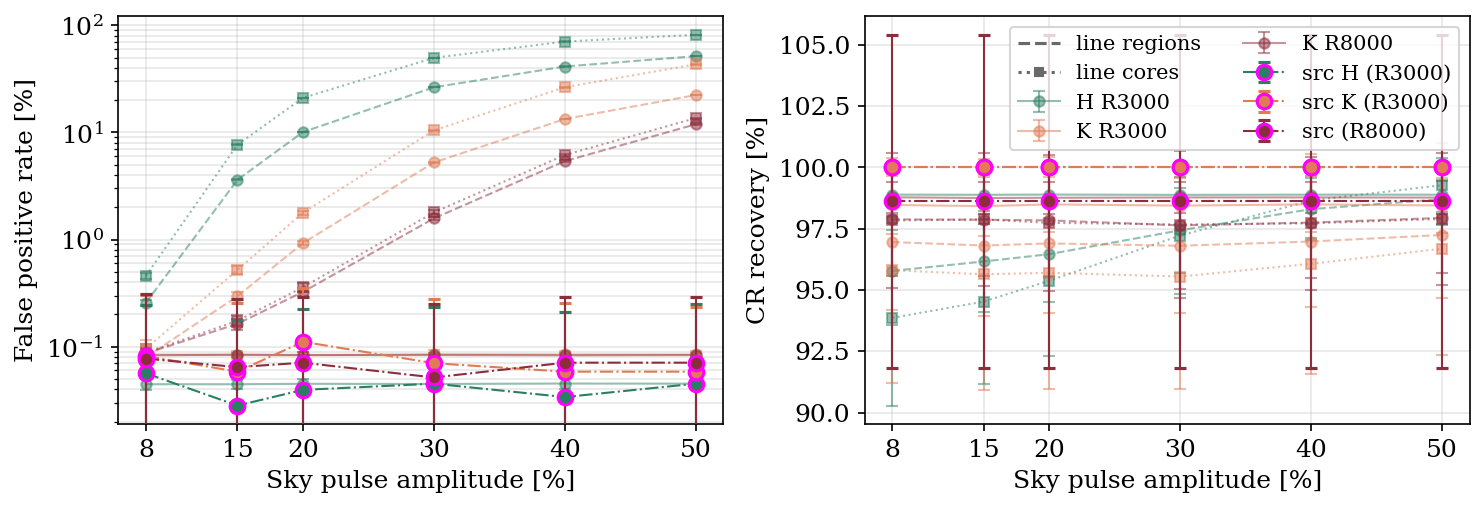}
\end{tabular}
\end{center}
\caption[example] 
{ \label{fig:sky_sweep} 
Mean false positive and CR recovery rates for a sweep of sky pulses between 8\% (the baseline case presented throughout this work) and 50\%, Means and standard deviations are calculated across 100 noise realizations. The sky linear drift is held constant to 2\%. Solid lines connect the inter-line region measurements while sky emission line regions are connected by dashed lines, sky emission line cores are plotted separately with squared markers. Measurements for source pixels are highlighted with magenta marker edges. Error bars correspond to the standard deviation across noise realizations.}
\end{figure}

To test the performance of the CR rejection under stronger sky variability we performed a sky amplitude sweep. Simulations were run for 100 noise realizations, for the \texttt{variable} sky case, for pulse amplitude percentages of 8, 15, 20, 30, 40, and 50\%, the time span of the pulses is fixed, as in the rest of this work, by the sigma of the Gaussian pulse, to 20 seconds, resulting in a total pulse duration of $\sim$60 s. The linear sky drift was held fixed to 2\% as in the previous simulations. False positive and CR recovery rates are presented in Figure \ref{fig:sky_sweep}. For sources and inter-line regions, the mean false positive rates remain below 0.1\%, except for the K-band R3000 source at 20\% pulse intensity, probably an artifact of the low number of pixels in the source mask. Sky emission lines increase the false positive rate as the pulse amplitude increases, reaching the maximum of $\sim$50\% for H-band R3000 sky lines at 50\% pulse amplitude.The 0.26\% FP rate over H-band sky lines from Table \ref{tab:cr_rejection} corresponds to the 8\% pulse amplitude point in the sweep, confirming consistency between the two independent sets of simulations.


Notably, the false positives are concentrated on the brightest emission line cores rather than distributed uniformly across the detector, demonstrating that the ramp fitter does not indiscriminately flag all pixels during a sky pulse

\subsection{Data volume estimates}

The adoption of the UTR readout carries a non-negligible impact on data volume. A single raw IFU file stores the full 4D data cube (NINT × NGROUPS × 2048 × 2048 pixels) as 16-bit integers, amounting to approximately 471 MB per file. For 7 hours of on-target plus sky exposures, accounting for science, sky, telluric star, flat, dark, Ronchi flat, and arc lamp frames in proportions that emulate the execution of an observing program, the four IFU detectors produce 382 GB of raw data per night; the Imager, assumed to operate in Fowler mode with 10 s exposures, contributes an additional 15 GB. Processed data products generated by DRAGONS, stored in 32-bit floating point with SCI, DQ, and VAR extensions, add 24 GB, yielding a total nightly data volume of approximately 421 GB. For the MCDS-8 readout mode, the expected data volume for the 4 IFUs is 14 GB per night, which, after adding the Imager and total reduced data yields a total of $\sim$ 54 GB per night, i.e. $\sim$ 8 times less data volume than in the UTR readout case. These estimates were computed using FITS files generated by the \texttt{GIRMOS Data Simulator} and should be taken into account when planning archive capacity and data transfer infrastructure for GIRMOS operations.

Although this represents an increase of roughly an order of magnitude in raw storage, the recovered on-sky integration time (Subsection \ref{sec:bias}) is, in operational terms, a scarcer resource than archive storage; we therefore consider the higher data volume an acceptable cost of the UTR readout.

\section{SUMMARY}

We have conducted a Monte Carlo experiment using the \texttt{GIRMOS Data Simulator} to assess whether the UTR sampling technique remains viable under variable sky conditions that break the assumed time-linearity of the signal and could cause false positive flags when detecting CRs. For this, detector images were simulated assuming three different cases: \texttt{no\_sky} (absent), \texttt{stable}, and \texttt{variable sky}. Our results show:

\begin{enumerate}
    \item For faint, read-noise-dominated targets, placed in the dark inter-sky-line regions, the UTR readout does not introduce a significant systematic bias in the recovered count rates relative to MCDS-8. The largest bias, of 0.15\% occurs over sky lines in the K-band for the R3000 resolution. This bias is negligible compared to typical flux calibration uncertainties of the NIR of $\sim$ 2-5\%.

    \item Sources located in inter-line regions show SNR changes consistent with the surrounding continuum. In the H-band R3000 mode, inter-line pixels show a median SNR improvement of 3-4\% for UTR over MCDS-8 under variable sky. In the K-band, the thermal spectrum increases the shot noise, degrading the SNR of continuum and source pixels by 2–3\% in the R3000 mode. The K-band R8000 mode presents an intermediate case, with bluer regions showing up to 2\% improvement while redder regions show 1–2\% degradation. Across all configurations, sky emission line pixels are shot-noise dominated, and UTR degrades their SNR by $\sim$2\%.

    \item CR recovery rates are independent of variability and always above 96\%, increasing to 98.5\% outside sky emission lines. For the GIRMOS baseline of 8\% sky pulse amplitude, FP rates show no sky-mode dependence except for H-band emission lines, where a FP rate of 0.25\% is found. The sky pulse amplitude sweep shows that FP rates over sky lines increase with pulse amplitude, reaching $\sim$ 80\% at 50\% amplitude at the line cores, but remain below 0.1\% for inter-line regions and source emission lines; CR recovery remains constant throughout the amplitude sweep. The partial flagging of sky-line pixels points to the interplay of deterministic sky jumps and stochastic shot-noise driving the false detections.

    \item Although the UTR readout increases the raw data volume saved to disk by roughly an order of magnitude relative to an on-instrument processed Fowler 2D frame, the SNR gain in some configurations translates into saved on-sky integration time, which is a more expensive resource than archive storage; we propose that the trade-off therefore favors UTR despite the higher data volume. Conversely, for configurations where the SNR is degraded, the benefit of UTR is less clear.
\end{enumerate}

These results use GSAOI readout noise values as a proxy; the actual GIRMOS detector characterization may require threshold recalibration. Additionally, as 1/f noise is not included in our simulations, its interaction with jump detection under variable sky will be better understood once detector data is available. The 0.15 $e^-$/s dark rate is a conservative early-laboratory value, a lower dark current would keep more pixels read-noise-limited and would increase the UTR advantage, so the improvements reported here can be considered, in that sense, as a lower bound.

The negligible bias observed in inter-line regions suggests that UTR fitting is robust for read-noise-limited targets under the sky variability levels adopted here (8\% peak-to-peak). This is a conservative estimate, as reported variability of up to 10\% corresponds to emission lines \cite{Ramsay1992}, while inter-line regions are expected to be considerably more stable. These results would be strengthened by systematic sky brightness measurements across multiple nights and at different times, to characterize how the sky varies near twilight on the timescales relevant for UTR fitting. Additionally, future work could explore the impact of variations in the seeing during an integration, which is assumed stable in the present simulations.

Our study supports the implementation of UTR readout for GIRMOS and similar read-noise-limited ground-based NIR spectrographs. The estimated SNR improvements for the H-band at R3000 will translate into savings of 3-4\% in valuable observing time with a manageable increase in data reduction software efforts and data archive volume. The performance of the UTR readout mode for the remaining GIRMOS configurations (J-band at R3000/R8000, H-band at R8000, and for the lower pixel scales) is expected to be higher than that reported here for the configurations most affected by thermal emission.

\section{APPENDIX}

\subsection{CR model}
\label{sec:cr_model}

The CR model to be applied in the simulations is of particular importance for testing whether sky jumps are wrongly flagged as CRs. A correct model must account for both a realistic shape in the CR hits and a realistic energy distribution with strong emphasis on representing low-energy CRs which are the most susceptible to being confused with sky variations. 

CRs typically present different particle composition, hit rate, and energy distributions depending on altitude and shielding. For instance, JWST measured a mean hit rate of 4.2 ions/cm$^2$/s\cite{Rauscher2025} with  CRs composed mainly of nucleons (H, He, C, N, O, Fe) \cite{Robberto2010}, while at sea level, the hit rate has been measured to be 0.014–0.016 hits/cm$^2$/s and composed mainly of muons, with a scale factor of 1.6 for typical $\sim$2700 m observatories and of 2.1 for Mauna Kea \cite{Groom2004}. Groom also reported different proportions and rates depending on shielding. Hence, it is of crucial importance to implement a CR model that represents the environment where GIRMOS will operate. GSAOI provides the best available empirical CR library for GIRMOS because it uses the same detector technology (H2RG HgCdTe) at a comparable high-altitude observatory site. The hit rate will need to be scaled for the altitude difference, but the morphology and energy spectrum of individual events are expected to be representative. We used GSAOI dark frames of 60s exposure from two observing runs (January and May 2013) to detect CRs above a 5$\sigma$ level exceeding a robust median-combined background after applying the GSAOI bad-pixel-mask. 

The selected candidates are then filtered by shape and charge level (e-). Candidates are filtered to have less than 50 pixels and to have an elongation $<$1.5 (axis ratio) if $N_{pix}>$ 6 to exclude rounded sources (more probably attributable to persistence or clustered bad pixels). The resulting library has 364,812 events obtained from the 25 dark frames (4 H2RG chips each) which translates to 4.47 hits/cm$^2$/s when accounting for exposure time and detector size. 

\begin{figure} [ht]
\begin{center}
\begin{tabular}{c} 
\includegraphics[height=9cm]{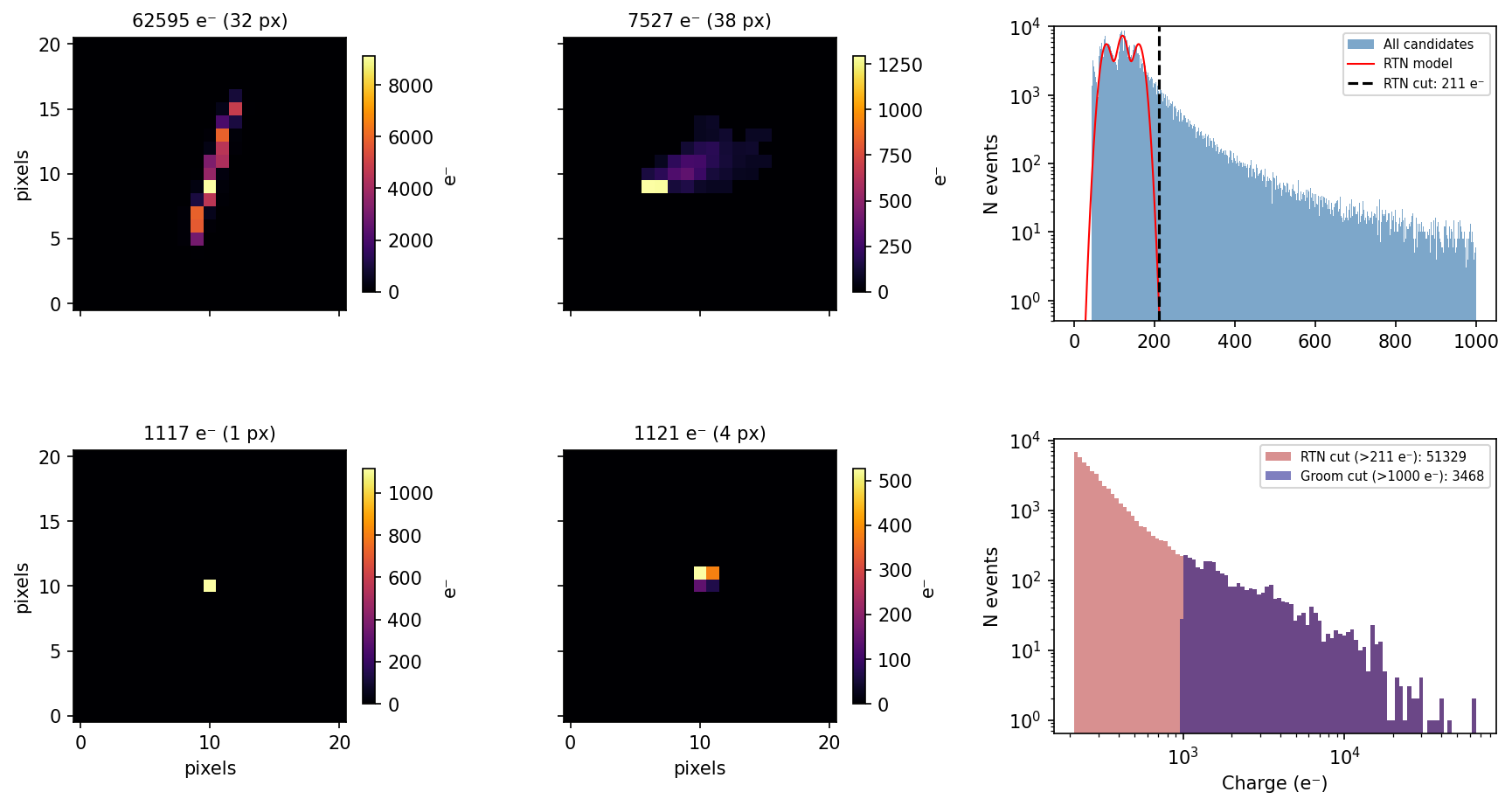}
\end{tabular}
\end{center}
\caption[example] 
{ \label{fig:cr_stamps} 
Left and middle panels: Examples of CR events that are common to the two constructed libraries. The cutouts have zero-values everywhere except where a CR was detected. Each pixel corresponding to a CR shows the empirically measured energy in $e^-$. Panel titles give the total charge and the number of affected pixels. Top right panel: Zoom-in into the lower energy distribution of all the events before cutting. The red curve corresponds to the RTN model fitted to the event energies. The black dashed line indicates the energy cut for the RTN library. Bottom right panel: Total energy distribution after energy cuts, the two libraries coincide in all the events with energies above 1000 e$^-$.}
\end{figure}

This hit rate is significantly higher than the rate expected for a ground-based telescope. A zoom in on the lower charge region of the total energy distribution of the candidates is depicted in the top right panel of Figure \ref{fig:cr_stamps}. These low energy detections are probably a mixture of detector artifacts, including CRs, but the multi-peak  shape suggests the presence of a large population of RTN (Random Telegraph Noise or Random Telegraph Signal, \cite{Kohley2018}) affected pixels. Following Ref. \cite{Bevidas2025} we fitted a 3-component symmetrical Gaussian model, with the constraints of the two flanking Gaussians having the same intensity and the same shift (different sign) w.r.t. the main Gaussian peak of the distribution, the fit is represented with a red curve in the figure. We generated two libraries: one excluding detections below the cutoff where the RTN model contribution drops below 1 event per bin, which corresponds to 211 e$^-$, and a second library with an energy cut of 1000 e$^-$ that gives a rate consistent with the expected CR flux at Mauna Kea's altitude \cite{Groom2004}. For these two libraries we obtained hit rates of 0.63 and 0.043 hits/cm$^2$/s respectively. The lower rate (0.043) is consistent with expected ground-based CR flux, while the higher rate (0.63) likely retains significant RTN contamination, providing an upper bound on the CR population. A sample of CRs is shown in Figure \ref{fig:cr_stamps} along with the energy distributions of the CRs in each library, dubbed RTN and Groom respectively, the legend includes the total number of events in each library. 

\acknowledgments 
We thank K. Labrie, C. Simpson, and W. Vacca from the Gemini/DRAGONS team for discussions that helped shape this investigation.
This work was supported by funding from the Canada Foundation for Innovation and the Nova Scotia Research and Innovation Trust.

\bibliography{report} 
\bibliographystyle{spiebib} 

\end{document}